\documentclass[preprintnumbers,superscriptaddress,showpacs,pra]{revtex4}
\usepackage{epsfig}
\usepackage{graphicx}
\usepackage{color}

\input{tcilatex}
\begin{document}

\title{A simple realization of $gl(2,c)$ Lie algebra with vector operators
on sphere}
\author{Q. H. Liu}
\affiliation{School for Theoretical Physics, and Department of Applied Physics, Hunan
University, Changsha, 410082, China}
\author{X. P. Rong}
\affiliation{School for Theoretical Physics, and Department of Applied Physics, Hunan
University, Changsha, 410082, China}
\author{D. M. Xun}
\affiliation{School for Theoretical Physics, and Department of Applied Physics, Hunan
University, Changsha, 410082, China}
\date{\today}

\begin{abstract}
By utilization of three elementary vector operators as position, angular
momentum and their cross product, a simple realization of $gl(2,c)$ Lie
algebra on sphere are constructed. The coherent states based on this algebra
can then be constructed by standard manner.
\end{abstract}

\maketitle

\section{Introduction}

Position operator $\mathbf{R}$, angular momentum operator $\boldsymbol{L}$
and their cross product as $\mathbf{R}\times \boldsymbol{L}$ are three
typical vector operators \cite{1,2} that are widely used to explore the
quantum properties of the angular momentum \cite{3}, and various coherent
states on the spheres \cite{4,5,6,7}. However, in contrast with the former
two operators that are well-studied, the fundamental properties of the
operator $\mathbf{R}\times \boldsymbol{L}$ are still not fully understoond
yet. In this paper, we present an algebraic analysis of the linear
combination of three vector operators of following form, 
\begin{equation}
\mathbf{J=}\frac{i}{\hbar }\mathbf{N}\times \mathbf{L+}a\mathbf{N+}\frac{b}{%
\hbar }\mathbf{N}L_{z},  \label{1}
\end{equation}%
where $\mathbf{N}\boldsymbol{\equiv }\mathbf{R/}R$ is the unit directional
vector, and $a$ and $b$ are two parameters. We are going to point out that,
with proper combinations, this vector (\ref{1}) together with $\boldsymbol{L}
$ can be used to form the $gl(2,c)$ Lie algebra. As well-known, the $gl(2,c)$
algebra is in general the four dimensional Lie algebra $G(a_{0},b_{0})$ with 
$a_{0}\ $and $b_{0}$ being two parameters, composed of operators $K_{+}$, $%
K_{-}$, and $K_{z}$ and $I$; and these operators satisfy commutation
relations \cite{8}:\ 
\begin{equation}
\left\{ 
\begin{array}{c}
\lbrack K_{+},K_{-}]=2a_{0}K_{z}+b_{0}I \\ 
\lbrack K_{z},K_{\pm }]=\pm K_{\pm } \\ 
\lbrack \mathbf{K},I]=0%
\end{array}%
\right. ,  \label{2}
\end{equation}%
where $I$ is a $c$-number, $\mathbf{K=(}K_{x},K_{y},K_{z}\mathbf{)}$ and $%
K_{\pm }=K_{x}\pm iK_{y}$ as usual. Two Casimir operators are \cite{8}:

\begin{equation}
\left\{ 
\begin{array}{c}
C=K_{+}K_{-}+a_{0}K_{z}^{2}-\left( a_{0}+b_{0}\right) K_{z} \\ 
C_{0}=I%
\end{array}%
\right. .  \label{3}
\end{equation}%
As $a_{0}\ $and $b_{0}$ take values $0$ or $\pm 1$, one has the following
algebras \cite{8,9,10,11}:%
\begin{eqnarray}
G(1,\pm 1) &:&o(3)\oplus u(1)\approx u(2)\oplus u(1),  \label{4} \\
G(-1,\pm 1) &:&o(2,1)\oplus u(1)\approx u(1,1)\oplus u(1),  \label{5} \\
G(1,0) &:&so(3)\oplus u(1)\approx su(2)\oplus u(1),  \label{6} \\
G(-1,0) &:&so(2,1)\oplus u(1)\approx su(1,1)\oplus u(1),  \label{7}
\end{eqnarray}%
and etc \cite{8,9,10,11}.

This paper is organized as what follows. In section II, the raising operator 
$K_{+}$ and the lowing operator $K_{-}$ are formed by the proper
construction of vectors $\mathbf{J}$ (\ref{1}). In section III, an algebraic
analysis is performed. In last section IV, a brief conclusion is given.

\section{Raising operator $K_{+}$ and lowing operator $K_{-}$}

For convenience, we write down two operators of similar form, 
\begin{equation}
\left\{ 
\begin{array}{c}
\mathbf{J}_{1}\mathbf{=}\frac{i}{\hbar }\mathbf{N}\times \mathbf{L+}a\mathbf{%
N+}\frac{b}{\hbar }\mathbf{N}L_{z} \\ 
\mathbf{J}_{2}\mathbf{=-}\frac{i}{\hbar }\mathbf{N}\times \mathbf{L+}c%
\mathbf{N+}\frac{d}{\hbar }\mathbf{N}L_{z}%
\end{array}%
\right. ,  \label{8}
\end{equation}%
where $c$ and $d$ are new parameters similar to $a$ and $b$. The hermitian
conjugate of these two operators $\mathbf{J}_{1}$\ and $\mathbf{J}_{2}$ are
given by respectively,%
\begin{equation}
\left\{ 
\begin{array}{c}
\mathbf{J}_{1}^{\dagger }=-\frac{i}{\hbar }\mathbf{N}\times \mathbf{L+}%
a^{\ast }\mathbf{N+}\frac{b^{\ast }}{\hbar }\mathbf{N}L_{z}-2\mathbf{N+}%
\frac{b^{\ast }}{\hbar }[L_{z},\mathbf{N}] \\ 
\mathbf{J}_{2}^{\dagger }=\frac{i}{\hbar }\mathbf{N}\times \mathbf{L+}%
c^{\ast }\mathbf{N+}\frac{d^{\ast }}{\hbar }\mathbf{N}L_{z}+2\mathbf{N+}%
\frac{d^{\ast }}{\hbar }[L_{z},\mathbf{N}]%
\end{array}%
\right. .  \label{9}
\end{equation}%
Now, we define raising operator $K_{+}$ and lowing operator $K_{-}$ by $%
K_{+}\equiv J_{1x}+iJ_{1y}$\ and $K_{-}\equiv J_{2x}-iJ_{2y}$, 
\begin{eqnarray}
K_{+} &\mathbf{=}&\frac{i}{\hbar }(\mathbf{N}\times \mathbf{L)}_{+}\mathbf{+}%
aN_{+}+bN_{+}\frac{L_{z}}{\hbar },  \label{10} \\
K_{-} &\mathbf{=}&\mathbf{-}\frac{i}{\hbar }(\mathbf{N}\times \mathbf{L)}_{-}%
\mathbf{+}cN_{-}+dN_{-}\frac{L_{z}}{\hbar }.  \label{11}
\end{eqnarray}%
Their hermitian conjugates are after calculations, 
\begin{equation}
K_{+}^{\dagger }=\mathbf{-}\frac{i}{\hbar }(\mathbf{N}\times \mathbf{L)}%
_{-}+(a^{\ast }-2-b^{\ast })N_{-}+b^{\ast }N_{-}\frac{L_{z}}{\hbar },
\label{12}
\end{equation}%
\begin{equation}
K_{-}^{\dagger }=\frac{i}{\hbar }(\mathbf{N}\times \mathbf{L)}_{+}+(c^{\ast
}+2+d^{\ast })N_{+}+d^{\ast }N_{+}\frac{L_{z}}{\hbar }.  \label{13}
\end{equation}%
We impose following requirement $K_{+}^{\dagger }=K_{-}$ on these operators,
and find, 
\begin{equation}
\left\{ 
\begin{array}{c}
a^{\ast }-2-b^{\ast }=c \\ 
b^{\ast }=d%
\end{array}%
\right. .  \label{14}
\end{equation}%
Taking the hermitian conjugate twice, we will get back to the same operator $%
K_{-}^{\dagger }=(K_{+}^{\dagger })^{\dag }=K_{+}$; and the results
identical to Eq. (\ref{14}) will be obtained. For further simplification, we
require that these parameters $a$, $b$, $c$ and $d$ are all real numbers. In
consequence, there are only two parameters are independent, which can be
conveniently chosen as $a$ and $b$, 
\begin{equation}
c=a-b-2,d=b.  \label{15}
\end{equation}%
Finally we get the the raising and the lowing operator $K_{+}$ and $K_{-}$
respectively, 
\begin{equation}
K_{+}\mathbf{=}\frac{i}{\hbar }(\mathbf{N}\times \mathbf{L)}_{+}\mathbf{+}%
aN_{+}+bN_{+}\frac{L_{z}}{\hbar },\text{and }K_{-}\mathbf{=}\mathbf{-}\frac{i%
}{\hbar }(\mathbf{N}\times \mathbf{L)}_{-}\mathbf{+}(a-b-2)N_{-}+bN_{-}\frac{%
L_{z}}{\hbar }.  \label{16}
\end{equation}%
The third operator $K_{z}$ is not known yet, which will be determined by
requiring that that three operators $\mathbf{(}K_{x},K_{y},K_{z}\mathbf{)}$
form the $gl(2,c)$ Lie algebra through the communication relations (\ref{2}%
). The explicit form of $K_{z}$ will be carried out in following section.

\section{Lie algebra analysis of $K_{+}$ and $K_{-}$}

Direct computation can give following relations,%
\begin{equation}
\lbrack
K_{+},K_{-}]=-(a+c)(b+1)-((a+c)(1-b))N_{z}^{2}-2(1+2b+b^{2}+b(1-b)N_{z}^{2})%
\frac{L_{z}}{\hbar },  \label{17}
\end{equation}%
\begin{equation}
\lbrack L_{z}/\hbar ,K_{\pm }]=\pm K_{\pm }.  \label{18}
\end{equation}%
These two relations suffice to meet our need, though there are still other
communication relations. The closure of the $gl(2,c)$ Lie algebra implies
that two coefficients before $N_{z}^{2}$ are zero,$\ $%
\begin{equation}
\left\{ 
\begin{array}{c}
(a+c)(1-b)=0 \\ 
b(1-b)=0%
\end{array}%
\right. .  \label{19}
\end{equation}%
Thus, with redefining operator $K_{z}\equiv L_{z}/\hbar $, we finally obtain
the $gl(2,c)$ Lie algebra: 
\begin{equation}
\lbrack K_{+},K_{-}]=-(a+c)(b+1)-2(b+1)^{2}K_{z},\text{ }[K_{z},K_{\pm
}]=\pm K_{\pm }.  \label{20}
\end{equation}%
All possibilities of the independent parameters $a$ and $b$ in Eq. (\ref{20}%
) are given by association of Eqs. (\ref{15}) and (\ref{19}) 
\begin{equation}
\left\{ 
\begin{array}{l}
i)\text{ }b=0,a=1 \\ 
ii)\text{ }b=1,a\neq 1 \\ 
iii)\text{ }b=1,a=1%
\end{array}%
\right. .  \label{21}
\end{equation}%
So, we have only three decompositions of the $gl(2,c)$ Lie algebra:\ Case 1, 
$su(1,1)$ with $a=-c=1\ $and $b=d=0$, case 2, $u(1,1)\otimes $ $u(1)$ with $%
b=d=1$ and $a+c\neq 0$, and finally case 3, $su(1,1)$ with $b=d=1$ and $%
a=-c=1$. In all these cases, $K_{z}\equiv L_{z}/\hbar $, and for case 1, 2,
and 3, communication relations $[K_{+},K_{-}]$ are given by, respectively, 
\begin{equation}
\left\{ 
\begin{array}{c}
K_{x}=i\left( \frac{i}{\hbar }\mathbf{N}\times \mathbf{L+N}\right) _{y} \\ 
K_{y}=-i\left( \frac{i}{\hbar }\mathbf{N}\times \mathbf{L+N}\right) _{x}%
\end{array}%
\right. ,
\end{equation}%
\begin{equation}
\left\{ 
\begin{array}{c}
K_{x}=\left( \frac{(a+c)}{2}\mathbf{N+}\frac{1}{\hbar }\mathbf{N}%
L_{z}\right) _{x}+i\left( \frac{i}{\hbar }\mathbf{N}\times \mathbf{L+}\frac{%
(a-c)}{2}\mathbf{N}\right) _{y} \\ 
K_{y}=-i\left\{ \left( \frac{i}{\hbar }\mathbf{N}\times \mathbf{L+}\frac{%
(a-c)}{2}\mathbf{N}\right) _{x}+i\left( \frac{(a+c)}{2}\mathbf{N+}\frac{1}{%
\hbar }\mathbf{N}L_{z}\right) _{y}\right\}%
\end{array}%
\right. ,
\end{equation}%
\begin{equation}
\left\{ 
\begin{array}{c}
K_{x}=\left( \frac{1}{\hbar }\mathbf{N}L_{z}\right) _{x}+i\left( \frac{i}{%
\hbar }\mathbf{N}\times \mathbf{L+N}\right) _{y} \\ 
K_{y}=-i\left( \frac{i}{\hbar }\mathbf{N}\times \mathbf{L+N}\right)
_{x}+\left( \frac{1}{\hbar }\mathbf{N}L_{z}\right) _{y}%
\end{array}%
\right. .
\end{equation}

It is easy to see that the Casimir operator $C$ (\ref{3}) does not commute
with the Hamiltonian for free particles constrained on the sphere, so it do
not represent any invariant for the particles.

\section{Conclusion and Discussion}

After analyzing vector operators constructed by directional and angular
momentum operator, we find that they can form Lie algebras of type $gl(2,c)$
that can be decomposed into three cases but two algebraic types as $%
u(1,1)\otimes $ $u(1)$ and $su(1,1)$. The unitary irreducible representation
for each of these algebras $u(1,1)\otimes $ $u(1)$ and $su(1,1)$ can be
obtained by standard manner \cite{8,9,10,11}. Also from the manner, we can
construct the coherent states based on these algebras \cite{5,6,7,8}, but
the relation between these states and the motion of the particle constrained
on the sphere is an intricate problem \cite{4,5,6,7}. We raise but leave
this issue open for further debate and exploration.

\end{document}